\documentclass[12pt, letterpaper]{article}

\pdfoutput=1

\usepackage{color}

\usepackage{amsmath, amssymb, graphics, epsfig, graphicx, epstopdf}

\usepackage{amsfonts}
\usepackage{amssymb}
\usepackage{graphicx}
\usepackage{amstext}
\usepackage{color}

\newcommand{\nc}{\newcommand}
\nc{\ba}{\begin{eqnarray}}
\nc{\ea}{\end{eqnarray}}

\newcommand{\bea}{\begin{eqnarray}}
\newcommand{\eea}{\end{eqnarray}}

\nc{\be}{\begin{eqnarray}}
\nc{\ee}{\end{eqnarray}}

\nc{\bfk}{{\bf k }}
\nc{\bfx}{{\bf x }}
\nc{\pfp}{{\bf{p}}}
\nc{\bfp}{{\bf{p}}}
\nc{\bfq}{{\bf{q}}}
\nc{\tbf}{\textbf}

\nc{\calP}{  { \cal P} }
\nc{\calR}{  { \cal R} }
\nc{\im}{ \mathrm{Im} }
\nc{\sg}{ \mathrm{sgn} }
\begin{document}

\vspace{5mm}
\vspace{0.5cm}
\begin{center}

\def\thefootnote{\fnsymbol{footnote}}

{\Large   Primordial inhomogeneities from  massive defects during inflation}
\\[0.5cm]

{  Hassan Firouzjahi,  Asieh Karami, Tahereh Rostami  }
\\[0.3cm]
{\small \textit{School of Astronomy, Institute for Research in Fundamental Sciences (IPM) \\ P.~O.~Box 19395-5531, Tehran, Iran
}}\\

\vspace{0.5cm}
{\small {e-mails:
 firouz@ipm.ir, ~ karami@ipm.ir, ~ t.rostami@ipm.ir
}}\\

\end{center}

\vspace{.8cm}

\hrule \vspace{0.5cm}


\begin{abstract}

We consider the imprints of local massive defects, such as a black hole or a massive monopole, during inflation. The massive defect breaks the background homogeneity. We consider the limit that the physical Schwarzschild radius of the defect is much smaller than the inflationary Hubble radius so
a perturbative analysis is allowed. The inhomogeneities induced in scalar and gravitational wave
 power spectrum are calculated. We obtain the amplitudes of dipole, quadrupole and octupole anisotropies in curvature perturbation power spectrum and identify the relative configuration of the defect to CMB sphere in which
large observable dipole asymmetry can be generated.
We observe a curious reflection symmetry  in which the configuration where the defect is inside the CMB comoving sphere has the same inhomogeneous variance  as its mirror configuration where the defect is outside the CMB sphere.

\end{abstract}
\vspace{0.5cm} \hrule
\def\thefootnote{\arabic{footnote}}
\setcounter{footnote}{0}
\newpage
\section{Introduction}

One of the original motivations for inflationary paradigm was to solve the monopole problem  \cite{Guth:1997wk, Guth:1980zm}. A rapid period of inflationary expansion dilutes all classical inhomogeneities
and defects such as monopoles and strings, providing a natural solution to  the problem of overproduction of primordial magnetic monopoles in models of grand unified theories  \cite{Guth:1979bh}.
With this picture in mind, there were not much attentions on defects during inflation. It is
natural to think that if inflation continues for long enough period, then the patch of inflationary background  encompassing the current observable Universe may have no monopole, justifying a simple isotropic and homogeneous inflationary patch to start with.

In this work we would like to study the effects of a local massive defect, such as a black hole or a monopole, during inflation.
If inflation does not last very long, i.e. not much longer than the 60 e-folds required to solve the flatness and the horizon problem, then it is likely that the existence of defects will have observational imprints on cosmological observations, specially on horizon size scales corresponding to low $\ell$ multipoles in CMB maps. Indeed, there are indications of deviations from  nearly scale-invariant and isotropic primordial power spectrum as predicted by simplest models of inflation on horizon scales  such as power deficit and hemispherical asymmetry \cite{Ade:2015lrj}. Because of the   cosmic variance the significance of these low-$\ell$ anomalies is under debate. However, if these anomalies have cosmological origins, then they hint towards more complicated dynamics of inflation with new degrees of freedom beyond the simple picture based on a slow-rolling scalar field.
With this motivation, the possible hemispherical asymmetry in CMB maps, as suggested  in WMAP and Planck data  \cite{Eriksen:2007pc, Ade:2013nlj, Ade:2015hxq},  has attracted significant interests in recent years, see also   \cite{Aiola:2015rqa, Mukherjee:2015mma, Mukherjee:2015wra, Adhikari:2014mua}) for recent data analysis on the search for dipole asymmetry in CMB maps.

There is no physically compelling mechanism to generate hemispherical asymmetry.  One intriguing proposal  is the mechanism of  long mode modulations  \cite{aniso-longmode}. In this approach
a  mode  which is much bigger than the Hubble radius generates the asymmetry by modulating the background inflationary parameters. However, it is well known that this proposal does not work in
simple single field models. This is because the amplitude of dipole modulation is related to the amplitude of local-type non-Gaussianity $f_{NL}$ as demonstrated in \cite{consistency}. 
Therefore, in  single field models of inflation with small (actually zero) $f_{NL}$ no dipole asymmetry is generated. This suggests one has to look  for models beyond the single slow roll setup  such as  curvaton scenarios or  iso-curvature perturbations, see \cite{various} for a list of various
theoretical works in generating observable dipole asymmetry.
In particular the proposal that a domain wall during inflation can be behind the observed dipole asymmetry was put forward in \cite{Jazayeri:2014nya}. It was shown that a scale-dependent large dipole can be generated in this setup while the amplitudes of higher multipoles such as quadrupole and octupole are small. This feature is particularly appealing, since the observations seem to prefer a scale-dependent dipole amplitude which falls off
on small CMB scales \cite{Ade:2013nlj, Ade:2015hxq, Akrami:2014eta}. These interesting results may
single out the roles of defects during inflation in addressing the
observed CMB anomalies.

In addition to \cite{Jazayeri:2014nya},  the fingerprints of primordial defects on curvature power spectrum have been studied in \cite{Carroll:2008br, Tseng:2009xw, Prokopec:2010nm, Wang:2011pb, Cho:2009en, Cho:2014nka}. In \cite{Tseng:2009xw} the correction to curvature perturbation power spectrum from
a  cosmic string during inflation is obtained. Since cosmic string breaks the rotational invariance, the  induced power spectrum breaks both rotational invariance and the translational invariance.

In this work, similar to the method employed in  \cite{Jazayeri:2014nya},  we calculate the corrections in curvature perturbation power spectrum from a massive defect during inflation. In addition, we calculate the corrections in gravitational wave power spectrum. Since the local mass term singularity breaks the translational invariance, our result  for power spectrum  maximally violates  the
translational invariance, i.e. there is no $\delta^3(\bfk + \bfq)$ in Fourier space for the modes
 $\bfk$ and $\bfq$.

The rest of the paper is organized as follows. In Section \ref{setup} we present our setup of
inflationary black holes and construct the required interaction Hamiltonian.
In Section \ref{power-sec}  we calculate the corrections in curvature perturbation power spectrum. In Section \ref{variance} the
variance of curvature perturbation power spectrum is calculated and  the amplitudes of dipole, quadrupole
and octupole in variance  are obtained. In Section \ref{GW} we calculate the corrections in gravitational wave power spectrum followed
by discussions in Section \ref{summary}. Some technical analysis for the interaction of tensor perturbations are relegated into the Appendix.


\section{Massive defect in Inflationary Backgrounds }
\label{setup}

In this section we present our set up. As mentioned before, we consider a local massive defect, i.e. a black hole, in inflationary background. In a sense, this massive defect may be viewed as a monopole too.
But technically  speaking a monopole is charged under the $U(1)$ gauge field. Therefore, our analysis
may not directly apply to a monopole. However, if one neglects the electromagnetic interactions of monopole and consider only its gravitational effects, then our results can be applied to monopole too.

Following the strategy employed in  \cite{Tseng:2009xw} and \cite{Jazayeri:2014nya},  in order to study the imprints of the massive defect on cosmological observations such as curvature perturbation power spectrum, we need to know the metric of the background in the presence of the defect. In the limit that one neglects the gravitational back-reactions of the inflaton field, the presence of the defect is felt by the inflaton field
via the deformation of the inflationary metric by the defect.  Happily the metric of local mass singularity
in a cosmological background is known. A spherically symmetric time-dependent solution of the Einstein equations that describes a black hole embedded
in an  FRW universe is  given by the McVittie
solution \cite{McVittie:1933zz, Kaloper:2010ec, Shiromizu:2001bg}
\ba
\label{planar}
ds^2 = - \left(1-\frac{GM}{2a(t)r}\right)^2
\left(1+\frac{GM}{2a(t)r}\right) ^{-2} dt^2 + a(t)^2 \left( 1+\frac{GM}{2a(t)r} \right)^4  d \vec{\bfx}^2 \, ,
\ea
where $r=|\vec \bfx|$ and $a(t)$ is the cosmic scale factor.  In this coordinate system black hole's physical  horizon corresponds to
$a(t) r=GM/2$ while the cosmological horizon is given by $H^{-1}$.
The  dS limit corresponds to the case in which $a(t) = e^{H t}$ with the Hubble expansion rate
$H$ being constant.  In this case, the hypersurface $GM/2a(t)r=1$ is regular as if the scalar curvature reduces to $R=12H^2$. It represents a space-like surface inside the event horizon \cite{Kaloper:2010ec}.

In order to have a consistent inflationary setup, we assume that the inflationary background is nearly a dS solution in which the dominant source of the energy density is provided by the inflaton field potential
$V(\phi)$ so the Hubble expansion rate is nearly given by $3 M_P^2 H^2 \simeq V(\phi)$ in which
$M_P^2=1/8 \pi G$ is the reduced Planck mass and
the  approximation holds in the slow-roll limit when we neglect the variation  of H, $- \dot H/H^2 \ll 1 $.
We assume that the massive defect is a small perturbation to the background slow-roll inflationary setup
so the energy sourced by the defect in a Hubble volume is much smaller than the inflaton potential. This corresponds to $M H^3 \ll M_P^2 H^2  $ or $G M H \ll 1$. In addition, in order to have a physically reliable analysis, we need to assume that the black hole's physical event horizon is much smaller than the cosmological horizon. Interestingly, this condition also requires $G M H \ll 1$. This discussion suggests that the dimensionless parameter $\mu \equiv  G M H$ is the key perturbative parameter in our analysis, in which the consistency of our assumptions requires $\mu < 1$.
In this approximation the interesting cosmological scales are far beyond the black hole's physical event horizon and we are justified to consider a power series  expansion of  the metric Eq. (\ref{planar}) in terms of $G M/a(t) r$.

With these discussions in mind, now let us look into the effects of the massive defect on inflaton's dynamics. As mentioned above, we neglect the back-reaction of inflaton on geometry so the background metric is given by Eq. (\ref{planar}) to leading order in slow-roll corrections. Then the presence of the massive defect is felt by the inflaton field mainly via deformation of the near dS metric by the defect as given in Eq. (\ref{planar}).

To be specific, we consider a massless scalar field  $\phi$ with the canonically normalized kinetic energy. The  leading contribution to curvature perturbation $\calR$ is given by $\calR = -H \delta \phi /\dot \phi$
in which $\delta \phi$ is the quantum fluctuations associated with the inflaton field. Now in order to
calculate the correction in curvature perturbation power spectrum we need to find the change in the
Hamiltonian of the inflaton field fluctuations $\delta \phi$  induced by the massive defect. This is given by the  second-order Lagrangian of  the scalar field fluctuations
\begin{equation}
\label{lag}
\mathcal L = \sqrt{-g}\left(-\frac{1}{2}g^{\mu\nu} \partial_{\mu} \delta \phi \partial_{\nu} \delta \phi \right),
\end{equation}
in which the metric is given by Eq. (\ref{planar}).  Expanding the metric to
leading orders in $GM/r a$ (considering the length scale much larger that the black hole's physical Schwarzschild radius)  the quadratic Lagrangian density is obtained to be
\begin{equation}
\label{lag1}
\mathcal L ={1\over 2}a^3\left[\left(1+\frac{4MG}{ra}+\frac{29M^2G^2}{4r^2a^2}\right)\delta \dot{\phi}^2-\left(1-\frac{M^2G^2}{4r^2a^2}\right)\frac{(\nabla \delta \phi)^2}{a^2}\right],
\end{equation}
in which a dot indicates the derivative with respect to cosmic time $t$ and $\nabla$ represents the usual
gradient with respect to the spatial comoving coordinate $\bfx_i$.

In order to use the standard in-in formalism to calculate the correction in power spectrum, we need the
interaction Hamiltonian. To do so  we  calculate the conjugate momentum associated with $\delta \dot \phi$, given by
\begin{equation}\label{Mom}
\Pi=a^3\left(1+\frac{4MG}{ra}+\frac{29M^2G^2}{4r^2a^2}\right){\delta \dot{\phi}}.
\end{equation}
Correspondingly, the quadratic Hamiltonian density is given by ${\cal H}={ {\cal H}}_0 + {{\cal H}}_I$ in which
${\cal H}_0 $ is the free Hamiltonian density, when $M=0$, given by
\ba
 \mathcal{H}_0=\frac{1}{2a^3}\Pi^2+ \frac{a}{2} (\nabla \delta \phi)^2  \, ,
\ea
while the interaction Hamiltonian density ${\cal H}_I$ to leading orders is given by
\ba
\label{inter}
{\cal H}_I= {1\over 2a^3}\left(-\frac{4MG}{ra}+\frac{35M^2G^2}{4r^2a^2}\right)\Pi^2-\frac{a}{2}\left(\frac{M^2G^2}{4r^2a^2}\right)(\nabla \delta \phi)^2 \, .
\ea
By replacing $\Pi$ with $\delta\dot{\phi}$ (with $\delta\dot{\phi}=\frac{\partial\mathcal{H}_0}{\partial\Pi}$ ),
we obtain the interaction Hamiltonian density in terms of $\delta \dot \phi$ which is more suitable for the
in-in analysis:
\be
\label{H-int}
\mathcal{H}_I={a^3\over 2} \left(-\frac{4MG}{ra}+\frac{35M^2G^2}{4r^2a^2}\right)\delta\dot{\phi}^2-
\frac{a}{2} \left(\frac{M^2G^2}{4r^2a^2}\right) (\nabla \delta \phi)^2 \, .
\ee
For the reasons which become clear soon, we have kept terms of quadratic order in $M^2 G^2$ in
the interaction Hamiltonian.

\section{Inhomogeneous Power Spectrum }
\label{power-sec}

Here we calculate the correction to curvature perturbation power spectrum. As mentioned before, we neglect the gravitational back-reaction of inflaton so the curvature perturbation is given by the usual formula
in flat gauge $\calR =-H \delta \phi/\dot \phi$.  We define the correction to the curvature perturbation power spectrum in Fourier space by
\be
\langle \calR_\bfk \calR_\bfq \rangle = \left( \frac{H}{\dot \phi}  \right)^2  \Big(\big\langle \delta \phi_\bfk \delta \phi_\bfq \big\rangle + \Delta\big\langle \delta \phi_\bfk \delta \phi_\bfq \big\rangle \Big),
\ee
where the first term is the standard, isotropic and homogenous power spectrum while the second term is the correction from the massive defect.

We calculate the correction to the power spectrum by taking
${H_I} = \int d^3 x {\cal H}_I$ as the leading interaction Hamiltonian. The power spectrum is calculated in Fourier space, therefore, we also need to calculate $H_I$ in Fourier space. The following  Fourier transforms are useful:
\begin{eqnarray}
  \frac{1}{|x|}&=&\frac{1}{2\pi^2}\int_{-\infty}^{\infty} \frac{d^3 \bfk}{|{\bfk}|^2}e^{i \bfk \cdot  \bfx} \\
  \frac{1}{|x|^2}&=&\frac{1}{4\pi}\int_{-\infty}^{\infty} \frac{d^3 \bfk}{|{\bfk}|}e^{i \bfk  \cdot \bfx} \, .
\end{eqnarray}

Let us first calculate the corrections to power spectrum to first order in $G M$. Using the above formulas, the leading order $H_I$ is given by
\be
\label{interaction}
H_I=-\frac{2MG}{(2\pi)^3(2\pi^2)}\int \frac{d^3\bfk d^3\bfq}{|{\bfk}+{\bfq}|^2}\delta \phi'(k) \delta \phi'(q)
\ee
in which a prime here and below denotes the derivative with respect to the conformal time  $\tau$
defined as usual via $d \tau = dt/a(t)$.

Using the standard in-in formalism, the correction in power spectrum induced from the
defect is obtained to be  \cite{Maldacena:2002vr, Weinberg:2005vy}
\ba
\label{delta-power0}
\Delta \big \langle \delta \phi_\bfk \delta \phi_\bfq \big \rangle  &=&
+i\int_{- \infty}^{t_e} \Big \langle \big[H_I(\tau), \delta \phi_{\bfk} \delta \phi_{\bfq} \big] \Big\rangle d t
\\
&=&
\frac{2\mu}{H^2(2\pi)^3(2\pi^2)}\int_{-\infty}^{\tau_e}\frac{d\tau}{\tau} \int\frac{d^3\bfk'd^3\bfq'}{|\bfk'+\bfq'|^2} \im \Big[ \Big \langle \delta \phi'_{\bfq'} (\tau) \delta \phi'_{\bfk'}(\tau)  \delta \phi_\bfk(\tau_e) \delta \phi_\bfq(\tau_e)  \Big \rangle
\Big] \, , \nonumber
\ea
in which $\tau_e$ represents the value of conformal time  at the end of inflation, $\tau_e \rightarrow 0$.
We see that this corrections in power spectrum is linear in terms of the dimensionless
parameter $\mu$.

 Calculating the expectation values using the Wick's theorem, we obtain
 \be
 \label{delta-power2a}
\Delta  \big \langle \delta \phi_\bfk \delta \phi_\bfq \big \rangle  = - \frac{16\pi \mu}{H^2}\frac{1}{|\bfk+\bfq|^2}\int\frac{d\tau}{\tau}
\im \Big[ \delta \phi'_{q}(\tau) \delta \phi'_{k}(\tau) \delta \phi_q^*(\tau_e) \delta \phi_k^*(\tau_e)\Big]+
 k \leftrightarrow q.
 \ee
 The wave function of the free theory is given by
 \be
 \delta \phi_k = \frac{H}{\sqrt{2 k^3}} (1+ i k \tau) e^{-i k \tau}.
 \ee
 Plugging this into the integral in Eq.~(\ref{delta-power2a}) and using $\tau_e \simeq 0$, the term containing $\im [ ...] $ in Eq.~(\ref{delta-power2a}) becomes proportional to
  $$I \equiv \int_{-\infty} ^0 d \tau \tau  e^{-i ( k + q) \tau} \quad. $$
Using the contour rotation $\tau = -\infty (1+ i \mu_0)$ with $\mu_0 \rightarrow 0^+$, the UV contribution in $I$ is canceled, while the IR contribution from $\tau =0$ in $I$ is found to be  real. As a result, $\im (I)=0$. Therefore, to first order in $\mu$ there is no correction in curvature perturbation power spectrum. This is consistent with the result obtained in \cite{Cho:2014nka}.  This means that in order to calculate the corrections in power spectrum, we have to go to second order in $\mu$, i.e. consider the interaction Hamiltonian containing the term $(GM)^2$. This was the reason why we have kept terms quadratic in $GM$ in the interaction Hamiltonian in Eq. (\ref{H-int}).

Now we calculate the corrections to  curvature power spectrum to second order in $\mu$.
There are three different types of contributions to  order $\mu^2$. The first two corrections come from the direct contribution of the terms containing $(G M)^2$ from operators $(\delta \dot \phi)^2$ and $(\nabla \delta \phi)^2$
in Eq. (\ref{H-int}) into the in-in integral Eq. (\ref{delta-power0}).  The third contribution comes from the
product of two $H_I$  linear in $\mu$, as given in Eq. (\ref{interaction}),  in  a nested integral as we elaborate in details below.
But, first we concentrate on the direction contributions of operators $(\delta \dot \phi)^2$ and $(\nabla \delta \phi)^2$ quadratic in $\mu$.

The Hamiltonian in Fourier space for the interaction $(\delta \dot \phi)^2$ is given by
\ba
\label{H-a}
 H_I^{(a)}=\frac{35M^2G^2}{8(2\pi)^3(4\pi) a(t)}\int \frac{d^3\bfk d^3\bfq}{|\bfk+\bfq|}\delta \phi'(k) \delta \phi'(q),
\ea
while the Hamiltonian for the operator $(\nabla \delta \phi)^2$ is given by
\ba
\label{H-b}
H_I^{(b)}=\frac{ a(t)M^2G^2}{8(2\pi)^3(4\pi)}\int \frac{d^3\bfk d^3\bfq(\bfk.\bfq)}{|\bfk+\bfq|}\delta \phi(k) \delta \phi(q) \, .
\ea

Let us first calculate the corrections from $ H_I^{(a)}$. Using $ H_I^{(a)}$ in the in-in integral in
Eq. (\ref{delta-power0}) we obtain
\be
\label{delta-power1}
\Delta \big \langle \delta \phi_\bfk \delta \phi_\bfq \big \rangle^{(a)}=\frac{35\mu^2\pi^2H^2}{4}\frac{1}{|\bfk+\bfq|}\frac{1}{kq}\im{\left[\int_{-\infty}^{\tau_e} d\tau\tau^2 e^{-i(k+q)\tau}\right]},
\ee
which yields
\be
\label{delta-power1}
\Delta \big \langle \delta \phi_\bfk \delta \phi_\bfq \big \rangle^{(a)}=-\frac{35\mu^2\pi^2H^2}{ 2}\frac{1}{|\bfk+\bfq|}\frac{1}{kq}\frac{1}{(k+q)^3} \, .
\ee

Similarly,  the corrections from the interaction $ H_I^{(b)}$ is given by
\be
\label{delta-power1}
\Delta \big \langle \delta \phi_\bfk \delta \phi_\bfq \big \rangle^{(b)}=\frac{\mu^2\pi^2H^2}{4}\frac{1}{|\bfk+\bfq|}\frac{\bfk.\bfq}{k^3q^3}\im {\left[\int_{-\infty}^{\tau_e} d\tau\left(1+i\tau(k+q)-kq\tau^2\right) e^{-i(k+q)\tau}\right]},
\ee
which leads to
\be
\label{delta-power2}
\Delta \big \langle \delta \phi_\bfk \delta \phi_\bfq \big \rangle^{(b)}=-\frac{\mu^2\pi^2H^2}{2}\frac{1}{|\bfk+\bfq|}\frac{\bfk.\bfq}{k^3q^3}\left[\frac{1}{(k+q)}+\frac{kq}{(k+q)^3}\right] \, .
\ee

Correspondingly, the total contribution from the interactions $ H_I^{(a)}$  and $ H_I^{(b)}$ is  given by
\be
\label{delta-power3}
\Delta \big \langle \delta \phi_\bfk \delta \phi_\bfq \big \rangle^{(a+b)}=-\frac{\mu^2\pi^2H^2}{2}\frac{1}{|\bfk+\bfq|}\left[\frac{35}{kq(k+q)^3} +\frac{\bfk.\bfq}{k^3q^3(k+q)}+ \frac{\bfk.\bfq}{k^2q^2(k+q)^3}\right] \, .
\ee

Finally, we calculate the third contributions, the products of two $H_I$ linear in $\mu$ in a nested
in-in integral. Similar to integrals in  \cite{Emami:2013bk,Chen:2014eua,   Akhshik:2014gja} this contribution is given by
\ba
\Delta \big \langle \delta \phi_\bfk \delta \phi_\bfq \big \rangle^{(c)}= - \int_{-\infty}^{\tau_e} d \tau_1
\int_{-\infty}^{\tau_1} d \tau_2 \Big \langle \Big[ H_I (\tau_2) , \Big[  H_I (\tau_1) , \delta \phi_\bfk(\tau_e)  \delta \phi_\bfq(\tau_e) \,  \Big]  \Big]\,
\Big \rangle \, ,
\ea
in which $H_I$ in the above integral is given by Eq. (\ref{interaction}), yielding
\ba
\Delta \big \langle \delta \phi_\bfk \delta \phi_\bfq \big \rangle^{(c)} = 16 (2 \pi)^3\,  \left(\frac{G M}{\pi^2}\right)^2 \int d^3 \bfp \frac{1}{| \bfp - \bfq|^2}  \frac{1}{| \bfp + \bfk|^2} \int_{-\infty}^{\tau_e} d\tau_1  a(\tau_1) \int_{-\infty}^{\tau_1} d\tau_2  a(\tau_2)  \nonumber\\
~~~~~~~~~~~~~~~~~~~~~~\times \im \Big[   \delta \phi_q(\tau_e)^* \delta \phi_q'(\tau_1)  \Big]
\im \Big[ \delta \phi_k(\tau_e)^* \delta \phi_k'(\tau_2)  \delta \phi_p'(\tau_1)^* \delta \phi_p'(\tau_2)
\Big]  + (k\rightarrow q) \, .
\ea
Note that the factor $16$ comes from various permutations in Wick contractions of  $\delta \phi$.
The additional integration over the momentum $\bfp$ is because of the violation of translation invariance
so we lose the usual $\delta^3(\bfk + \bfq)$ contribution which appears in homogeneous backgrounds.

The above integral over momentum $\bfp$ has non-trivial forms. For the UV contributions, i.e. when
$| k \tau|, |q\tau|,    | p \tau| \gg 1$, the integrands oscillate rapidly and the net result is zero \cite{Chen:2009zp}. Therefore, the dominant contributions in the integrals above come from the IR region, yielding
\ba
\Delta \big \langle \delta \phi_\bfk \delta \phi_\bfq \big \rangle^{(c)} = \frac{32 \mu^2}{\pi} H^2
\int d^3 \bfp \frac{1}{| \bfp - \bfq|^2}  \frac{1}{| \bfp + \bfk|^2} \frac{p ( 4 k^3+ 3 k^2 p + p q^2)}{k (k+ q)^3 (k+p)^2 (k-q)^3} + (k\rightarrow q) \, .
\ea

 Now changing $\bfp \rightarrow -\bfp$ will take care of the contribution from the permutation $k \rightarrow q$, so  the final result is
 \ba
 \label{delta-c}
 \Delta \big \langle \delta \phi_\bfk \delta \phi_\bfq \big \rangle^{(c)} = \frac{128}{\pi} \mu^2 H^2
\int d^3 \bfp \frac{1}{| \bfp - \bfq|^2}  \frac{1}{| \bfp + \bfk|^2} \frac{p^2 \big(  p^2 + 2 (k+q) p + (k^2 + q^2 + 3 kq)   \big)}{ k q (p+ q)^2 (p+ k)^2 (k+ q)^3  } .
 \ea
Despite the complicated integral over $\bfp$, the overall scaling of the above term is similar
to the other two contributions in
$\Delta \big \langle \delta \phi_\bfk \delta \phi_\bfq \big \rangle^{(a+b)}$, i.e. scaling like $\mu^2 H^2$.

In conclusion,  the total corrections in inhomogeneous power spectrum is given by  adding
$\Delta \big \langle \delta \phi_\bfk \delta \phi_\bfq \big \rangle^{(a+b)} $ and $  \Delta \big \langle \delta \phi_\bfk \delta \phi_\bfq \big \rangle^{(c)} $ obtained in Eqs.  (\ref{delta-power3}) and (\ref{delta-c}). Correspondingly, the power spectrum of curvature perturbations is obtained to be
\ba
\label{power-total}
\langle \calR_\bfk \calR_\bfq \rangle = \big( \frac{H^2}{\dot \phi} \big)^2  \Big\{ \frac{1}{2 k^3 } ( 2 \pi)^3
\delta^3(\bfk + \bfq) - \frac{\mu^2\pi^2}{2}\frac{1}{|\bfk+\bfq|}\Big[\frac{35}{kq(k+q)^3} +
\frac{\bfk.\bfq \left( k^2 + q^2 + 3 k q \right)}{k^3q^3(k+q)^3}\Big]
\nonumber\\
+ \frac{128}{\pi} \mu^2
\int d^3 \bfp \frac{1}{| \bfp - \bfq|^2}  \frac{1}{| \bfp + \bfk|^2} \frac{p^2 \big(  p^2 + 2 (k+q) p + (k^2 + q^2 + 3 kq)   \big)}{ k q (p+ q)^2 (p+ k)^2 (k+ q)^3  }
\Big\}
\ea
The first term above represents the dominant isotropic and homogeneous contribution coming from the
inflation field. As mentioned before, the contributions of the massive defect is at the order $\mu^2$ with a non-trivial scale-dependence. Since the homogeneity of the background is maximally broken
by the local defect, there is no $\delta^3(\bfk + \bfq)$ for the contribution of defect.
However, since the background is still isotropic, the correction in power spectrum  respects isotropy.

\section{Variance}
\label{variance}

Having calculated the corrections in curvature perturbation power spectrum, here we calculate
the variance of curvature perturbation in real space $\langle \calR(\bfx)^2 \rangle$ on CMB sphere. As argued in  \cite{Akrami:2014eta, Jazayeri:2014nya} the asymmetry in variance  in real space is a very good measure of the dipole asymmetry in power spectrum. Indeed, the asymmetry in variance of the temperature map (which is directly related to $\langle \calR(\bfx)^2 \rangle$ up to a numerical factor ) is introduced as
one of the measure of dipole asymmetry in latest Planck's data analysis \cite{Ade:2015hxq}.
Therefore, it is a very good idea to calculate $\langle \calR(\bfx)^2 \rangle$ for our setup.  A similar analysis was performed for the anisotropy induced from the domain wall in \cite{Jazayeri:2014nya}.

The variance has two parts, the leading homogeneous part coming from the inflaton fluctuations denoted by  $\langle \calR(\bfx)^2 \rangle^{(0)}$,   and the sub-leading correction from the  massive defect. The leading contribution  is related to  the curvature perturbation  powers spectrum  ${\calP_\calR}^{(0)}$ via
\ba
\langle \calR(\bfx)^2 \rangle^{(0)} = \int d \ln k\, \left(  \frac{k^3}{2 \pi^2} | \calR(\bfk)|^2 \right)  = \int d \ln k \, {\calP_\calR}^{(0)} \, .
\ea
On the other hand, the correction to $\langle \calR(\bfx)^2 \rangle$ from the massive defect  is given by
\ba
\Delta \langle \calR(\bfx)^2 \rangle = \frac{1}{(2 \pi)^6} \int\int\textrm{d}^3\bfk\textrm{d}^3\bfq  e^{i (\bfk+ \bfq)\cdot \bfx} \Delta \langle \calR_{\bfk} \calR_{\bfq} \rangle \, .
\ea
Using Eq. (\ref{power-total}), this yields
\ba
\label{var-final}
\Delta \langle \calR(\bfx)^2 \rangle &=&
 \left(\frac{ \mu H^2}{ (2 \pi)^3 \dot\phi}\right)^2  \int\textrm{d}^3\bfk\textrm{d}^3\bfq  e^{i (\bfk+ \bfq)\cdot \bfx}    \Bigg\{ \frac{- \pi^2}{2 k q |\bfk+\bfq| (k+q)^3}
 \Big(35+ \frac{\bfk.\bfq \left( k^2 + q^2 + 3 k q \right)}{k^2q^2}  \Big) \nonumber\\
&+&  ~~~~~ \frac{128}{\pi}
\int d^3 \bfp \frac{1}{| \bfp - \bfq|^2}  \frac{1}{| \bfp + \bfk|^2} \frac{p^2 \Big(  p^2 + 2 (k+q) p + (k^2 + q^2 + 3 kq)   \Big)}{ k q (p+ q)^2 (p+ k)^2 (k+ q)^3  }
\Bigg\} \, .
\ea

Obviously the above integrals are too complicated to be dealt with analytically. However, as in \cite{Jazayeri:2014nya},  important insights can be obtained by looking at the UV and IR properties of these integrals.  One can easily check that the UV part converges and plays no important roles in the integrals while the  important contributions  come from the IR region. One can check that upon rescaling $k\rightarrow rk$ , $q\rightarrow rq, p \rightarrow rp$, in which $r = | \bfx|$,  the integrand becomes independent of $r$ and any dependence on $r$ comes from the regularization of the IR cut-off for $k,q, p\rightarrow 0$.
Concentrating on the IR region of the integrals, one can check that the final result is in the form of
$\ln(r/L)$ in which $L$ is the size of a hypothetical  box which is assumed to be much bigger than the comoving size of the patch encompassing the observed CMB sphere.
With this analytical insight, we have fitted numerically the above integrals  with  the ansatz $\ln(r/L)$ and obtained
\ba
\label{avariance}
\Delta \langle \calR(\bfx)^2 \rangle \simeq
668\frac{\mu^2H^4}{4(2\pi)^4\dot\phi^2}\ln\Big(\frac{r}{L}\Big) + C_0 \, .
\ea
Here $C_0$ is a constant independent of $r$ which modifies the monopole but does not  contribute to higher multipoles.  We have checked that Eq. (\ref{avariance}) is a good fit to the full
numerical results of the integrals in Eq. (\ref{var-final}). Furthermore, we have checked numerically  that the dominant term in  $\Delta \langle \calR(\bfx)^2 \rangle$ is the third integral in   Eq. (\ref{var-final}). Specifically,  the amplitudes of the first and second integral ( without considering their coefficients) are around $2\pi^2$ but the amplitude of the
third integral is approximately $104\pi$.

\begin{figure}
  \hspace{0cm}
  \includegraphics[width=0.9\linewidth]{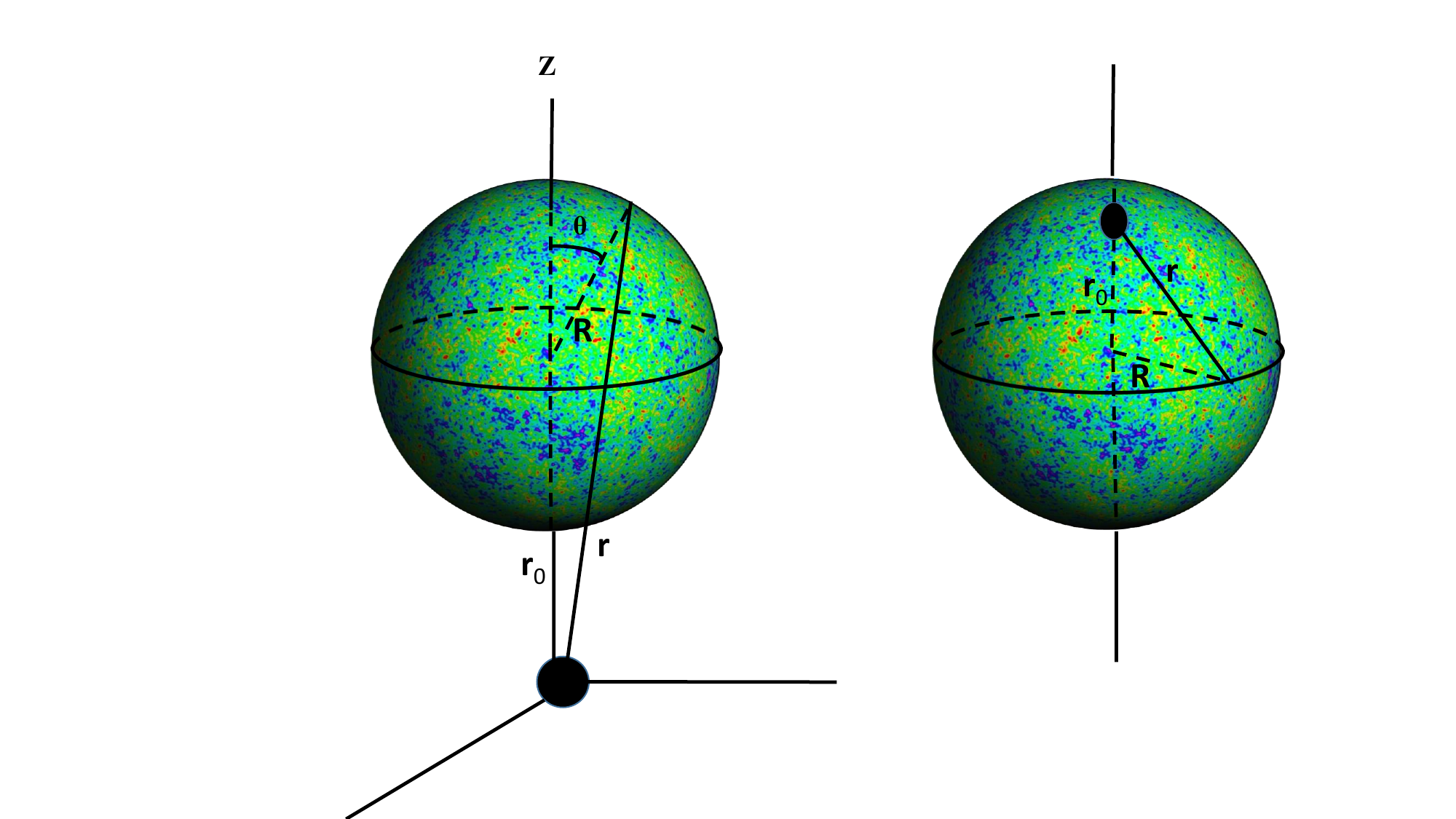}
  \caption{The setup for CMB sphere and massive defect. The defect is at the center of coordinate system and
  its distance from the center of CMB sphere is denoted by $r_0$.
  \textbf{Left:} The case where the massive defect is outside the CMB sphere. \textbf{Right:} The massive defect is inside the CMB sphere.}
  \label{cmbf}
\end{figure}

To obtain a measure of the CMB dipole and higher multipole asymmetries,  we consider a two-dimensional sphere which is  fixed at a  comoving radius $R$ centered at $\mathbf r_{\mathrm {CMB}}=r_0\mathbf{\hat z}$ as the CMB sphere. Because of the rotational symmetry, we can choose the $z$ axis to be the line connecting the center of CMB sphere to the massive defect (the origin).
For a view of this configuration see Fig. \ref{cmbf}. If $r_0 > R$, then the defect is outside the CMB sphere while the defect will be inside the CMB sphere when $r_0 <R$.

 The center of this CMB sphere is located
at comoving distance $r_0$ from the position of the monopole while any point on the CMB sphere is identified with two angles $\theta$ and  $\phi$. Because of the azimuthal symmetry,  the latter does not play any role and we have
\ba
r= |\bfx| = R \sqrt{1+ \alpha^2 - 2 \alpha \cos \theta } \, ,
\ea
in which $\alpha \equiv r_0/R$. Plugging this into Eq. (\ref{avariance}) we obtain
\be
\label{cmbvariance}
\Delta\langle\mathcal R^2({\bf r})\rangle\approx\mu^2\frac{668}{8(2\pi)^2}\mathcal P_0\ln\left(\alpha^2+1-2\alpha\cos\theta  \right)+C_0  \, ,
\ee
where $\mathcal P_0=(H^2/2\pi\dot\phi)^2$ is the homogeneous power spectrum.
Here we comment that in order for our perturbative analysis to be correct, we require that
the corrections in variance to be smaller than the isotropic and homogeneous one, i.e.
$\Delta\langle\mathcal R^2({\bf r})\rangle \ll \langle\mathcal R^2({\bf r})\rangle^{(0)} $.  Assuming that
the logarithmic term is not hierarchically much different than unity, this requires
$10\,  \mu^2 \ll 1$ which is well consistent with our approximation in which $\mu \ll 1$.

To calculate the dipole and higher multipoles for the variance of the
curvature perturbations, we decompose $\Delta\langle\mathcal R^2({\bf r})\rangle$ in terms of the Legendre polynomials as
\ba
\Delta\langle\mathcal R^2({\bf r})\rangle=\mathcal P_0\sum_\ell a_\ell P_\ell(\cos\theta) \, .
\ea
Correspondingly, the multipoles $a_\ell$ for $\ell \ge 1$ (i.e. neglecting the monopole which contains the unknown parameters $C_0$), are given by
\ba
a_\ell &=& \frac{2 \ell +1}{2} \int_{-1}^1 d (\cos\theta)  \Delta\langle\mathcal R^2({\bf r})\rangle P_\ell (\cos \theta)  \nonumber\\
\label{al-eq}
&\propto& \frac{2 \ell +1}{2} \int_{-1}^1 d (\cos\theta)   P_\ell (\cos \theta)  \ln\left(\alpha^2+1-2\alpha\cos\theta \right) \, .
\ea

\begin{figure}
\centering
\begin{minipage}{.45\textwidth}
  \centering
  \includegraphics[width=1.14\linewidth]{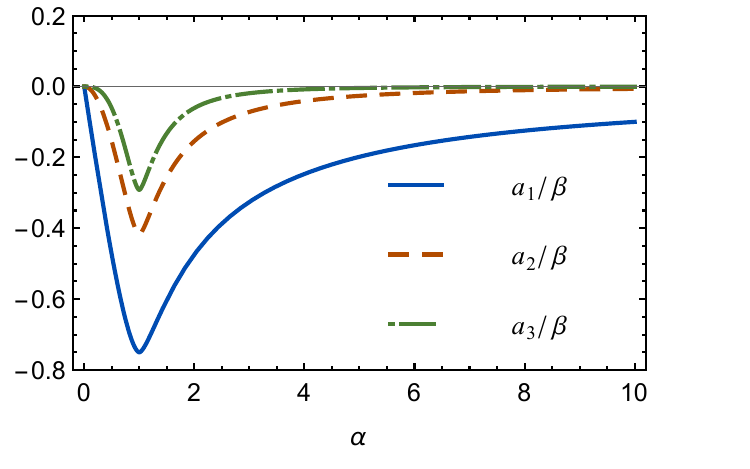}
  \label{dqo}
\end{minipage}%
\hspace{0.7cm}
\begin{minipage}{0.45\textwidth}
  \centering
  \includegraphics[width=1.14\linewidth]{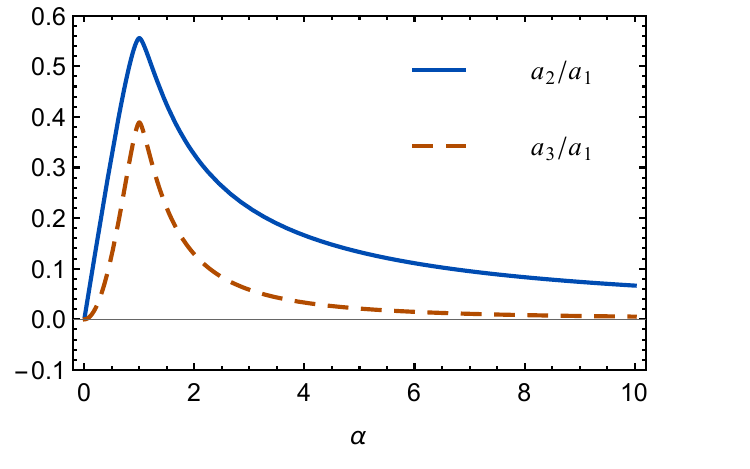}
  \label{dqoratio}
\end{minipage}
\caption{{\bf Left.} Values of the variance dipole ($a_1$), quadrupole ($a_2$) and octupole ($a_3$), scaled by $1/\beta$,
as functions of $\alpha$. {\bf Right.} The ratios of the quadrupole ($a_2$) and octupole ($a_3$) to the dipole ($a_1$) as functions of $\alpha$.}
\label{dqo-m}
\end{figure}

Combining  this decomposition with Eq. (\ref{cmbvariance}),  the dipole ($a_1$), quadrupole ($a_2$) and octupole ($a_3$) are obtained to be
\begin{eqnarray}
\label{poles}
a_1&=&\frac{ 3 \beta }{16\alpha^2}\Big[(\alpha^2-1)^2 \ln\Big|\frac{1+\alpha}{1-\alpha}\Big|-2\alpha(1+\alpha^2)\Big] \, ,\\
a_2&=&\frac{ 5 \beta }{96\alpha^3}\Big[ 3(\alpha^2-1)^2 (1+\alpha^2)\ln\Big|\frac{1+\alpha}{1-\alpha}\Big|-2\alpha(3-2\alpha^2+3\alpha^4)\Big] \, ,\\
a_3&=&\frac{ 7 \beta }{768 \alpha^4}\Big[3(\alpha^2-1)^2(5+6\alpha^2+5\alpha^4) \ln\Big|\frac{1+\alpha}{1-\alpha}\Big|-2 \alpha(15-7\alpha^2-7\alpha^4+15\alpha^6)\Big].
\end{eqnarray}
in which we have defined $\beta\equiv 668\mu^2/4(2\pi)^2$. As discussed before, the consistency of
our setup requires $\mu <1$ so we require $\beta <1$.

One interesting feature of the above results is that $a_i$ are symmetric under $\alpha \rightarrow 1/\alpha$. This has interesting interpretation. Suppose $r_0 > R$ so the massive defect is outside the CMB sphere and $\alpha >1$. Now consider a situation in which $\alpha \rightarrow 1/\alpha$ so the defect
is inside the CMB sphere with $r_0^{\mathrm{new}} = 1/r_0$. Then the variance for any point on the CMB sphere
remains unchanged.  This reflection symmetry can be verified from Eq. (\ref{al-eq}) for all values of
$\ell \ge 1$. Indeed, upon changing $\alpha \rightarrow 1/\alpha$, the right hand side of  Eq. (\ref{al-eq}) yields
\ba
 a_\ell - \ln (\alpha) \frac{2 \ell +1}{2}   \int_{-1}^1 d (\cos\theta)   P_\ell (\cos \theta)  \, .
\ea
But for $\ell \ge 1$ the integral above vanishes so we conclude $a_\ell \rightarrow a_\ell$ upon $\alpha \rightarrow 1/\alpha$.

One can check that $a_i$ reaches its maximum value when $\alpha=1$, i.e.
$r_0= R$ and the massive defect is located right on the surface of CMB sphere during inflation.  For small   values of  $\alpha$, one can check that
\ba
a_1 \simeq -\beta \alpha   \quad , \quad a_2 \simeq -\frac{2}{3} \beta \alpha^2 \quad , \quad
a_3 \simeq -\frac{8}{15} \beta \alpha^2  \quad , \quad (\alpha \ll 1 ) \, .
\ea

Finally, we comment that observations indicate a dipole amplitude at the order of few percents
\cite{Ade:2015hxq, Akrami:2014eta} while detecting no higher multipoles. The amplitude of dipole (and other multipoles) is proportional to the parameter $\beta=668\mu^2/4(2\pi)^2 $. As mentioned before, the consistency of our setup requires $\mu <1$. As an example, if we take  $\mu =1/10$, then we
obtain $\beta \simeq 0.04  $ so a dipole at the order of few percents can be obtained for $\alpha\sim 1$, i.e. when the defect is near the surface of CMB sphere. As can be seen from Fig. \ref{dqo-m} a large dipole and small other multipoles can be obtained for $\alpha \sim 1$. For $\alpha \ll 1$ ( or $\alpha \gg 1)$ the higher multipoles fall off rapidly. But the problem with these configurations  is that   for
these values of $\alpha$, dipole also falls off. So the configuration in which  the defect is somewhat near the CMB sphere,  either from the outside or from the inside,
is the preferred configuration observationally.

\section{Gravitational Waves}
\label{GW}

The presence of the massive defects also contributes into the gravitational waves power spectrum.
In this section we calculate the modification in tensor perturbation power spectrum.


The metric perturbations for tensor modes are given by
\ba
\label{hij}
ds^2 = - \left(1-\frac{GM}{2a(t)r}\right)^2
\left(1+\frac{GM}{2a(t)r}\right) ^{-2} dt^2 + a^2\left({1+\frac{MG}{2a(t)\, r}} \right)^4\left(\delta_{ij}+h_{ij}\right) dx^i \, dx^j \, ,
\ea
in which $h_{ij}$ represents the tensor perturbations. In addition, we fix the gauge freedom by using the transverse and traceless (TT) gauge: $ h_{ii}=0$ and $\partial_i  h_{ij}=0$. As usual this leaves two degrees of freedom for the tensor modes. Note that the indices on $h_{ij}$ are raised and lowered
by the flat metric $\delta_{ij}$.

As in the case of scalar perturbations, we have to calculate the interaction Hamiltonian for tensor perturbations. These interactions come from the Einstein-Hilbert term. The details of the analysis are
presented in the Appendix.

There are four types of interaction Hamiltonians for tensor perturbations as follows:
\be
\label{InterH}
&&H^{(1)}_I=-2\mu\pi{M_P^2\over H}\int d^3\mathbf{x}\,  \delta(r)(h_{ij})^2,\\
&&H^{(2)}_I=-2\mu {M_P^2\over H}\int d^3\mathbf{x} \, \partial_j\partial_k\left(\frac{1}{r}\right)h_{ij}h_{ik},\\
&&H^{(3)}_I=\frac{9}{8}\mu {M_P^2 H a^2 }\int d^3\mathbf{x}\frac{1}{r}h_{ij}^2,\\
\label{InterH4}
&&H^{(4)}_I=\mu {M_P^2 a^2 \over 8 H}\int d^3\mathbf{x}\frac{1}{r} \dot{h}_{ij}^2.
\ee
We use the convention that all repeated indices are summed over (unless mentioned otherwise).

\subsection{Polarization bases}
\label{pol-base}

We use the following decomposition for the tensor perturbations in Fourier space
\be
h_{ij}(\mathbf k)=\sum_s h^s(\mathbf k)e^s_{ij}(\mathbf k) \, ,
\ee
in which $s$ represents the polarization. For linear polarization $s=1,2$,  while for circular polarization, $s=+,  \times$,  with the following properties for the polarization tensor
\be\label{prop}
k_i e^s_{ij}(\bfk) =0,~~~~~~e^r_{ij}(\bfk) e^{s\ast}_{ij}(\bfk) =\delta^{rs},~~~~~~~~e^s_{ii}(\bfk) =0 \, .
\ee
The leading homogenous and isotropic tensor power spectrum is given by
\be
\label{h-power0}
\langle h^r(\mathbf k,\tau) h^s(\mathbf q,\tau)\rangle^{(0)}=(2\pi)^3\delta^3(\mathbf{k}+\mathbf{q}) |h(k,\tau)|^2
\delta^{rs},
\ee
in which $h(k,\tau)$ is the wave function of the tensor perturbations
\be
h(k,\tau)=\frac{2H}{M_P\sqrt{2k^3}}(1+ik\tau)e^{-ik\tau} \, .
\ee

Since any two different vectors in three-dimensional space  are coplanar and determine a unique plane, without losing generality we choose vectors $\bfk$ and $\bfq$ to be in $y-z$ plane and assume that $\bfk$ is in $z$ direction, $\bfk=k(0,0,1)$ and $\bfq=q(0,\sin\psi,\cos\psi)$ in which  $\psi$ represents the angle between the vectors $\bfk$ and $\bfq$.

Using this convention,  the circular polarization matrices associated to vectors $\bfk$ and $\bfq$ are given by
\be
e_{ij}^{\times}(\bfk)=\frac{i}{\sqrt 2}
\begin{pmatrix}
	 0 & 1 & 0 \\
	 1 & 0 & 0\\
	 0 & 0 & 0
\end{pmatrix} \quad , \quad
e_{ij}^{+}(\bfk)=\frac{1}{\sqrt 2}
\begin{pmatrix}
	-1 & 0 & 0 \\
     0 & 1 & 0\\
	 0 & 0 & 0
\end{pmatrix}
\ee
and
\be
e_{ij}^{\times}(\bfq)=\frac{i}{\sqrt 2}
\begin{pmatrix}
	0 & \cos\psi & -\sin\psi \\
	\cos\psi & 0 & 0\\
	-\sin\psi & 0 & 0
\end{pmatrix}  , \quad
e_{ij}^{+}(\bfq)=\frac{1}{\sqrt 2}
\begin{pmatrix}
	-1 & 0 & 0 \\
	0 & \cos^2\psi & -\sin\psi\cos\psi\\
	0 &-\sin\psi\cos\psi  & \sin^2\psi
\end{pmatrix}
\ee
Using this representation, one can easily check that the following relations hold which will be used in the follow up analysis
\be
&& e_{ij}^{+\ast}(\bfk)e_{ij}^{+\ast}(\bfq)=\frac{1}{2}(\cos^2\psi+1)\, ,\\
&& e_{ij}^{\times\ast}(\bfk)e_{ij}^{\times\ast}(\bfq)=-\cos\psi \, ,\\
&& e_{ij}^{+\ast}(\bfk)e_{ij}^{\times\ast}(\bfq)=0 \, ,
\ee
 and
 \be
 && k_jq_ke_{ik}^{+\ast}(\bfk)e_{ij}^{{ {+}} \ast}(\bfq)=-\frac{kq}{2}\sin^2\psi\cos\psi \, , \\
 && k_jq_ke_{ik}^{\times\ast}(\bfk)e_{ij}^{\times\ast}(\bfq)=\frac{kq}{2}\sin^2\psi\, , \\
 && k_jq_ke_{ik}^{+\ast}(\bfk)e_{ij}^{\times\ast}(\bfq)=0 \, , \\
  && k_jq_ke_{ik}^{\times \ast}(\bfk)e_{ij}^{+\ast}(\bfq)=0 \, .
 \ee


\subsection{Power spectrum of gravitational waves}

Now we are ready to calculate the corrections in tensor power spectra. As mentioned before, we have four different types of interaction Hamiltonians.
The leading order correction in tensor power spectrum induced from the Hamiltonian $H_I^{(i)}$ with
$i=1, 2, 3, 4$ is given by
\be
\Delta \big \langle h^r(\mathbf k)h^s(\mathbf q) \big \rangle^{(i)}=i\int_{-\infty}^{t_e}dt \Big \langle\left[H_I^{(i)},h^r(\mathbf k)h^s(\mathbf q)\right] \Big \rangle=-2\,  \im \, \int_{-\infty}^{t_e}dt \big \langle H_I^{(1)},h^r(\mathbf k)h^s(\mathbf q) \big \rangle \, .
\ee
Below we calculate the contribute from each interaction separately.


\subsubsection{Contribution from $H_I^{(1)}$}
Let us start with $H_I^{(1)}$. In Fourier space we have
\be
H_I^{(1)}=- \frac{2 \pi \mu}{H} {M_P^2 }\int \frac{d^3\mathbf{p_1}}{(2\pi)^3}\frac{d^3\mathbf{p_2}}{(2\pi)^3} h_{ij}(\mathbf p_1)h_{ij}(\mathbf p_2) \, .
\ee
Using the relation
\be
\delta(r)=\frac{1}{(2\pi)^3}\int d^3\mathbf{k}e^{-i\mathbf{k.r}},
\ee
the correction in tensor power spectrum induced from $H_I^{(1)}$ is given by
\be
\Delta \big \langle h^r(\mathbf k)h^s(\mathbf q)\big \rangle^{(1)}=-\frac{32 \pi \mu H^2}{M_P^2}\frac{1}{k^3q^3}\im\int_{-\infty}^{0}\frac{d\tau}{\tau}
e^{-i(k+q)\tau}\left(1+i(k+q)\tau-kq\tau^2\right) e^{r\ast}_{ij}(\mathbf k)e^{s\ast}_{ij}(\mathbf q) .
\ee
Now using the relations
\be
&&\im\left[\int_{-\infty}^{0}\frac{d\tau}{\tau}e^{-i(k+q)\tau}\right]=-\frac{\pi}{2},\nonumber
\\
&&\im\left[\int_{-\infty}^{0}\frac{d\tau}{\tau}e^{-i(k+q)\tau}i(k+q)\tau\right]=0,\nonumber
\\
&&\im\left[\int_{-\infty}^{0}\frac{d\tau}{\tau}e^{-i(k+q)\tau}(kq)\tau^2\right]=0,
\ee
and noting that $e^{r\ast}_{ij}(\mathbf k)e^{s\ast}_{ij}(\mathbf q)$ is real,
the contribution from  $H_I^{(1)}$ is obtained to be
\be
\Delta \big \langle h^r(\mathbf k)h^s(\mathbf q) \big\rangle^{(1)}=\frac{16\pi^2\mu H^2}{M_P^2k^3q^3}e^{r\ast}_{ij}(\mathbf k)
e^{s\ast}_{ij}(\mathbf q) \, .
\ee
As expected,  since the defect breaks the background homogeneity,  there is  no $\delta^3(\bfk + \bfq)$ in the above expression while the isotropy is kept intact.

\subsubsection{Contribution from $H_I^{(2)}$}

Now we calculate the corrections from $H_I^{(2)}$.

In the Fourier space we have
\be
H^{(2)}_I=4\mu {M_P^2\over 2}\int \frac{1}{H}\frac{d^3\mathbf{q_1}}{(2\pi)^3}\frac{d^3\mathbf{q_2}}{(2\pi)^3}\frac{1}{2\pi^2}\left(\frac{(\mathbf{q_1}+\mathbf{q_2})_j
(\mathbf{q_1}+\mathbf{q_2})_k}{|\mathbf{q_1}
+\mathbf{q_2}|^2}\right)h_{ij}(\mathbf{q_1})h_{ik}(\mathbf{q_2}) \, ,
\ee
which, using Eq. (\ref{prop}),  leads to
\be
\Delta \large \langle h^r(\mathbf k)h^s(\mathbf q) \large \rangle^{(2)}=-\frac{32 \pi^2 \mu H^2}{M_p^2}\frac{q_\ell k_j}{k^3q^3|\bfk+\bfq|^2}e_{i \ell}^{r\ast}(\bfk)e_{ij}^{s\ast}(\bfq) \, .
\ee

\subsubsection{ Contribution from $ H_I^{(3)}$}

Similarly,  for the contribution from $ H_I^{(3)}$ we obtain
\be
\Delta \large \langle h^r(\mathbf k)h^s(\mathbf q) \large \rangle^{(3)}=
\frac{ 72 \pi  \mu H^2 e^{r\ast}_{ij}(\mathbf k)e^{s\ast}_{ij}(\mathbf q) }{M_P^2 |\mathbf{k}+\mathbf{q}|^2q^3k^3} \im\left[\int_{-\infty}^{0}\frac{d\tau}{\tau^3}\left(1+i(k+q)\tau-kq\tau^2\right)e^{-i(k+q)\tau}\right] \, .
\nonumber
\ee
Using the relation
\be
\im\left[\int_{-\infty}^{0}\frac{d\tau}{\tau^3}\left(1+i(k+q)\tau-kq\tau^2\right)e^{-i(k+q)\tau}\right] =-\frac{\pi}{4}\left(k^2+q^2\right),
\ee
we obtain
\be
\Delta \large \langle h^r(\mathbf k)h^s(\mathbf q)\large \rangle^{(3)}=-\frac{18 \pi^{2}\mu H^2}{M_P^2}\frac{\left(k^2+q^2 \right)}{|\mathbf{k}+\mathbf{q}|^2q^3k^3}e^{r\ast}_{ij}(\mathbf k)e^{s\ast}_{ij}(\mathbf q).
\ee

\subsubsection{Contribution from $H_I^{(4)}$}
Finally,  for the contribution from $H_I^{(4)}$ we have
\be
\Delta\large \langle h^r(\mathbf k)h^s(\mathbf q)\large \rangle^{(4)}=\frac{32\pi H^2}{M_P^2}\mu\frac{1}{|\mathbf{k}+\mathbf{q}|^2kq} \im\left[\int_{-\infty}^{0}d\tau e^{-i(k+q)\tau}\tau\right]e^{r\ast}_{ij}(\mathbf k)e^{s\ast}_{ij}(\mathbf q)=0 \, .
\ee

\subsection{Total corrections in tensor power spectrum }

Having obtained the contribution from each $H_I^{(i)}$ as presented above, we can calculate the total corrections
in tensor power spectrum. Using the relations between the polarization matrices listed at the end of  Section \ref{pol-base} we obtain
\be
&& \Delta\langle h^\times(\mathbf k)h^\times(\mathbf q)\rangle=-\frac{2\pi^2\mu H^2}{M_P^2k^3q^3}\cos\psi\left(8-9\frac{(k^2+q^2)}{|\bfk+\bfq|^2}\right)-\frac{32 \pi^2 \mu H^2}{M_P^2}\frac{\sin^2\psi}{k^2q^2|\bfk+\bfq|^2} \, ,\\
&& \Delta\langle h^+(\mathbf k)h^+(\mathbf q)\rangle=\frac{\pi^2\mu H^2}{M_P^2k^3q^3}(\cos^2\psi+1)\left(8-9\frac{(k^2+q^2)}{|\bfk+\bfq|^2}\right)+\frac{32 \pi^2\mu H^2}{ M_P^2}\frac{\sin^2\psi\cos\psi}{k^2q^2|\bfk+\bfq|^2} \, ,
\ee
while there is no mixing between $+$ and $\times$ modes.

By adding the above results,  the total inhomogenous correction to tensor power spectrum is obtained to be
\ba
\Delta_{\mathrm{total}} \Big \langle h(\mathbf k)h(\mathbf q) \Big \rangle =
- \frac{\pi^2\mu H^2}{M_P^2k^3q^3} \, ( 1- \cos \psi)^2 \,
\frac{ k^2 + q^2 + 32 kq + 16kq \cos \psi  }{k^2 + q^2 + 2 k q \cos \psi} \, ,
\ea
in which the relation $|\bfk+\bfq|^2 = k^2 + q^2 + 2 k q \cos \psi$ has ben used.
Note that $\cos \psi = \bfk \cdot \bfq /k q$ so the  correction in tensor power spectrum is
statistically isotropic as expected. However, the homogeneity is specifically broken, so unlike the leading power spectra given in Eq. (\ref{h-power0}), we have no additional factor $ ( 2 \pi)^2 \delta ^3 (\bfk + \bfq)$.

The overall scale dependence of correction in tensor power spectrum is similar to the scalar case
in which $\Delta \big \langle h(\mathbf k)h(\mathbf q) \big \rangle \sim  1/k^6$. However, the big difference compared to the scalar perturbation is that the correction in tensor power spectrum
is linear in $\mu$.

\subsection{Variance of Tensor perturbations}

As in the case of scalar perturbations, we can also calculate the corrections in variance of tensor perturbations in real space. Parallel to scalar perturbation, the change in the variance of the tensor perturbations is given by
\ba
\label{t-variance}
\Delta\langle h_{ij}(\mathbf r)h_{ij}(\mathbf r)\rangle&=&\sum_s\int\frac{\mathrm d^3k\mathrm d^3q}{2\pi^6}e_{ij}^s(\bfk)e_{ij}^s(\bfq)\langle h(\mathbf k)h(\mathbf q)\rangle e^{i(\bfk+\bfq).\mathbf r} \, .
\ea
The structure of integral is somewhat similar to the scalar case and is too complicated to be calculated analytically. However, as in the case of scalar perturbations, the dominant contributions come from the
IR region of the integrals. Checking the IR limit of the above integral, we found a double logarithm behavior for the IR divergence. Performing a numerical approximation for the integrals we have found
\be
\Delta\langle h_{ij}h_{ij}\rangle(\theta)\sim \frac{\mu H^2}{8\pi^4M_P^2}
\ln\left(\frac{L}{r_0} \right)\ln\left(1+\alpha^2-2\alpha\cos\theta \right) \, ,
\ee
in which, as before, $\alpha=r_0/R$, $\theta$ is the usual polar angle on the CMB sphere as denoted in
Fig. \ref{cmbf} and $L$ represents the size of the box.  The unknown parameter
$L$ can be absorbed in the isotropic and homogeneous tensor power spectrum so it does not appear
in multipole moments of tensor anisotropies (like dipole, quadrupole etc).

In order for our perturbative treatment to be consistent, we require that the corrections in tenor power spectrum to be smaller than the leading isotropic and homogeneous tensor power spectrum. In terms of
variance this requirement is translated into  $\Delta\langle h_{ij}h_{ij}\rangle(\theta) \ll \langle h_{ij}h_{ij}\rangle^{(0)}$ in which $\langle h_{ij}h_{ij}\rangle^{(0)}$ is the variance from the leading isotropic and homogeneous tensor perturbations obtained from Eq. (\ref{h-power0}). Considering a scale invariant stochastic tensor perturbation we
have
\ba
\langle h_{ij}h_{ij}\rangle^{(0)} \sim \frac{H^2}{\pi^2 M_P^2} \int_{r_0}^L \frac{d\, k}{k} \sim
\frac{H^2}{\pi^2 M_P^2} \ln \left( \frac{L}{r_0} \right)  \, .
\ea
 As a result, we obtain
\ba
\frac{\Delta\langle h_{ij}h_{ij}\rangle(\theta)}{\langle h_{ij}h_{ij}\rangle^{(0)}} \sim  \frac{\mu}{\pi^2} \ln\left(1+\alpha^2-2\alpha\cos\theta \right) \, .
\ea
Assuming the logarithmic contribution is not hierarchically different than unity, the consistency
of our perturbative treatment is well justified with our assumption $\mu \ll1$.

\section{Summary and Discussions}
\label{summary}

In this work we have studied the imprints of local massive defects such as a monopole or black hole during inflation. As mentioned before, a distribution of massive defects is quickly diluted during inflation.
Therefore, it seems reasonable to study a single defect in a comoving Hubble patch during inflation.
The presence of the local massive defect breaks the homogeneity of the cosmological background while keeping the isotropy intact.

We have calculated the inhomogeneities induced in curvature perturbation and gravitational wave power spectra. In our treatment the effects of the massive defect is felt by the inflaton field via the corrections of
defect to background geometry.  We work in the limit in which the back-reaction of the inflaton field on background geometry is neglected. This is justified in leading order where this approximations has error of ${\cal O}(\beta \sqrt{\epsilon})$ in which $\epsilon = -\dot H/H^2$ is the slow-roll parameter. Therefore, these corrections can be neglected in the limit of small enough values of $\epsilon$.

We have calculated the anisotropy multipoles such as dipole, quadrupole and octupole induced from primordial
inhomogeneities. We have found that quadrupole and octupole are always smaller than the dipole. This is encouraging, as the Planck data seems to suggest the existence of a dipole with no detection of
quadrupole and octupole.  We have argued that the configuration with $\alpha \sim 1$, i.e, when the
defect is somewhat near the surface of the CMB sphere either from the outside or from the inside,  is the preferred configuration observationally.
We have observed a curious mirror symmetry upon $\alpha \rightarrow 1/\alpha$ in which the configuration with the massive defect being inside the comoving CMB sphere is mapped to its mirror configuration, i.e. $r_0 \rightarrow 1/r_0$. We have shown that the inhomogeneous corrections for both of these mirror images are identical.

With the primordial inhomogeneities in curvature perturbations and gravitational wave power spectra
calculated here, it would be very interesting to perform a CMB data analysis and compare the predictions of our setup with the Planck data. It is an interesting question to see if the inflationary universe
with  local massive inhomogeneity is a better fit to CMB data. For example, it is open to see
whether this picture can generate an acceptable amount of dipole amplitude and at the same time resolve other anomalies on CMB map such as the power deficit on large scales. This is an interesting question which is beyond the scope of our current purely theoretical investigation. We would like to come back to this question in future.

In our phenomenological approach, we have not specified the origin and the fate of massive defect.
It might have been generated  from a phase transition during or before inflation. Whatever its origin, we need the defect to evaporate  during reheating  so the Universe starts its isotropic and  homogeneous history. We do not know the mechanism in which the defect evaporates.  Perhaps this is  entangled to the mechanism of reheating which drags the energy not only from the inflaton field but also from the defect. Related to this question one may wonder if the definition of curvature perturbation in flat gauge as $\calR = -H \delta \phi/\dot \phi$ is well-defined in the presence of defect. Perhaps the definition of curvature perturbation on  a flat three-dimensional surface in the presence of defect with the metric Eq. (\ref{planar}) is unclear. To justify our approximation in taking
$\calR = -H \delta \phi/\dot \phi$ we consider the idealized situation in which the process of reheating and the decay of the defect happen instantaneously. Therefore, one can safely define the curvature perturbation on flat slice at the time of end of inflation when the correlation functions are calculated
as we did above. 

There are couple of other directions in which the current work can be extended. One interesting question is the imprints of a charged monopole in which not only its mass $M$ but also its electric (magnetic) charge  $Q$ appears
in the metric. This is the Reissner-Nordstrom-deSitter (RNdS) solution. The structure of RNdS metric in cosmological coordinate is more complicated than Eq. (\ref{planar}) with multiple horizons. In addition, the requirement of evading the naked singularity imposes the constraint $Q \leq M$ \cite{Faraoni:2014nba}.
It is an interesting question to see what kind of inhomogeneities the combination of $M$ and $Q$ induce
on curvature perturbations and gravitational waves. Another interesting question is to consider a distribution (network) of massive defects which are being diluted at the early stage of inflation. In the weak field approximation which will be relevant to our study, one can consider the effects of defects by the
superposition of each defect without back reacting to each other. Mathematically, the metric will be the
superposition of metrics in the form of Eq. (\ref{planar}). While the defects are being diluted, they
leave their imprints to scalar and tensor power spectra. These inhomogeneities may be viewed as
the snapshot  for the local position of these massive defects during inflation.

\vspace{0.7cm}

{\bf Acknowledgments:}  We would like to thank J. T. Firouzjaee,  S. Jazayeri and M. Wise
for insightful discussions and comments.

 \vspace{0.3cm}
\appendix
\label{App1}

\section{Einstein-Hilbert action}
In this Appendix we present the details of the analysis yielding the tensor perturbations interaction Hamiltonians, Eqs. (\ref{InterH}) - (\ref{InterH4}).

In ADM decomposition, the metric with the tensor perturbations are given by
\label{GW1}
\be\label{GW1-1}
N=\left(\frac{1-MG/2ar}{1+MG/2ar}\right),~~~~~~N_i=0,~~~~~~g_{ij}=a^2({1+MG/2ar})^4\left(\delta_{ij}+h_{ij}\right),
\ee
which yields
$$
\sqrt{\det (g_{ij})}=a^3 ({1+MG/2ar})^6\left(1-\frac{1}{4}h_{ij}^2\right).
$$
We need to calculate the three terms of the Einstein-Hilbert action which is
\be\label{action}
{S_{EH}}={M_P^2\over 2}\int \sqrt{\det (g_{ij})} N \left(^{(3)}R+{1\over N^2}\left(E_{ij}E^{ij}-E^2\right)\right),
\ee
where $^{(3)}R$ is the three-dimensional Ricci scalar associated with the spatial metric $g_{ij}$ and
$E_{ij}$ is the extrinsic curvature
\ba
E_{ij} &=& {1\over 2}\left(\dot{g}_{ij}-\nabla_i N_j-\nabla_j N_i\right)
\nonumber\\
&=&a^2H({1+MG/2ar})^3\left[\left(1-MG/2ar\right)\left(\delta_{ij}+h_{ij}\right)+\frac{1}{2H}\left(1+MG/2ar\right)\dot{h}_{ij}\right] \, ,
\ea
in which $\nabla$ represents the covariant derivative associated with the metric $g_{ij}$. In addition we have
\ba
E^{ij}&=&a^{-4}({1+MG/2ar})^{-8}\left[E_{ij}-h_{il}E_{jl}-h_{jk}E_{ki}+h_{il}h_{kj}E_{kl}+h_{im}h_{ml}E_{jl}+h_{jl}h_{kl}E_{ki}\right] \nonumber
\ea
and
\ba
E\equiv g^{ij}E_{ij}=a^{-2}({1+MG/2ar})^{-4}\left[E_{ii}-h_{ij}E_{ij}+h_{ik}h_{kj}E_{ij}\right].
\ea
We calculate each term of Einstein-Hilbert action  separately.

\subsection{Term containing $^{(3)}R$ }
Defining the conformal transformation via
$$
g_{ij}=a^2(1+MG/2ar)^4\left[\delta_{ij}+h_{ij}\right]\equiv \Omega^2\tilde g_{ij},~~~~~\Omega \equiv a(1+MG/2ar)^2
$$
Ricci scalar is obtained to be   \cite{Wald:1984rg}
$$
^{(3)}R=\Omega^{-2}\left[^{(3)}\tilde R-4 \tilde g^{ij}\tilde\nabla_i\tilde\nabla_j\ln\Omega-2g^{ij}(\tilde\partial_i\ln\Omega)(\tilde\partial_j\ln\Omega)\right].
$$
Then
\be
\sqrt{\textrm{det}(g_{ij})}N{}^{(3)}R&=&\sqrt{\textrm{det}(g_{ij})}N\Omega^{-2}\left[{}^{(3)}\tilde R-4\tilde g^{ij}\tilde\nabla_i\tilde\nabla_j\ln\Omega-2\tilde g^{ij}(\tilde\partial_i\ln\Omega)(\tilde\partial_j\ln\Omega)\right] \nonumber\\
&=&N\Omega(1-\frac{1}{4}h_{ij}^2)\left[{}^{(3)}\tilde R-2\tilde g^{ij}
\bigg(2\tilde\partial_i\tilde\partial_j\Omega-2\tilde\Gamma_{ij}^k\tilde\partial_k\Omega-\frac{\tilde\partial_i\Omega\tilde\partial_j\Omega}{\Omega}\bigg)\right] \nonumber \\
&\simeq&N\Omega(1-\frac{1}{4}h_{ij}^2)\Big[{}^{(3)}\tilde R-2\frac{\delta_{ij}-h_{ij}+h_{ik}h_{jk}}{\Omega}
\big(2\tilde\partial_i\tilde\partial_j\Omega-2\tilde\Gamma_{ij}^k\tilde\partial_k\Omega-\frac{\tilde\partial_i\Omega\tilde\partial_j\Omega}{\Omega}\big)\Big]  \nonumber
\ee
Now with
\be
{}^{(3)}\tilde R=h_{ij;kk}h_{ij}-\frac{1}{2}h_{ij;k}h_{jk;i}+\frac{3}{4}(h_{ij;k})^2
\ee
for the first term in the Einstein-Hilbert action we obtain
\be
\begin{split}
\sqrt{\textrm{det}(g_{ij})}N{}^{(3)}R=&N\Omega\bigg(h_{ij;kk}h_{ij}-\frac{1}{2}h_{ij;k}h_{jk;i}+\frac{3}{4}(h_{ij;k})^2\bigg)\\
&+N(h_{ij})^2\tilde\partial_k\tilde\partial_k\Omega-4Nh_{il}h_{lj}\tilde\partial_i\tilde\partial_j\Omega-8Nh_{ij}h_{ik;j}\tilde\partial_k\Omega\\
&+4Nh_{ij}h_{ij;k}\tilde\partial_k\Omega+\mathcal O(h^3).
\end{split}
\ee

Doing some integrations by part for the third term on the right hand side of the above equation,
the ${}^{(3)}R$ contribution to the Einstein-Hilbert action to first order in $\mu$ is obtained to be
\be
\label{first0}
 (\sqrt{\textrm{det}(g_{ij})}N{}^{(3)}R)^{(1)}=4\pi\frac{\mu}{H}\delta(r)(h_{ij})^2+4\frac{\mu}{H}\partial_j\partial_k\left(\frac{1}{r}\right)h_{ij}h_{ik} \, .
\ee

\subsubsection{Term containing $E_{ij}E^{ij}$ }

For this contribution we have
\be
E_{ij}E^{ij}=\frac{1}{a^4(1+MG/2ar)^{8}}\left[E_{ij}^2-2E_{ij}h_{il}E_{jl}+E_{ij}E_{kl}h_{il}h_{jk}+2E_{ij}h_{lj}h_{kl}E_{ki}\right] \, .
\ee
We calculate each of the above four terms in turn:
\be\label{first}
\begin{split}
  \sqrt{\textrm{det}(g_{ij})}\frac{N^{-1} E_{ij}^2}{a^4(1+MG/2ar)^{8}}  &=H^2 a^3\Bigg[\frac{1}{4}\left(1+2MG/ar\right)h_{ij}^2+\frac{1}{4H^2}\left(1+4MG/ar\right)\dot{h}_{ij}^2\\
  &+\frac{1}{H}\left(1+3MG/ar\right)\dot{h}_{ij}h_{ij}\Bigg],
\end{split}
\ee
\be\label{2}
\sqrt{\textrm{det}(g_{ij})}\frac{N^{-1} \left(-2E_{ij}h_{il}E_{lj}\right)}{a^4(1+MG/2ar)^{8}}=-4H^2 a^3\left[\left(1+2MG/ar\right)h_{ij}^2+\frac{1}{2H}\left(1+3MG/ar\right)\dot{h}_{ij}h_{ij}\right].
\ee
\be\label{3}
\sqrt{\textrm{det}(g_{ij})}\frac{N^{-1}}{a^4(1+MG/2ar)^{8}}\left(E_{ij}h_{il}h_{jk}E_{kl}\right)=H^2 a^3\left[\left(1+2MG/ar\right)h_{ij}^2\right],
\ee
\be\label{4}
\sqrt{\textrm{det}(g_{ij})}\frac{N^{-1}}{a^4(1+MG/2ar)^{8}}\left(2E_{ij}h_{jl}h_{lk}E_{ki}\right)=2H^2 a^3\left[\left(1+2MG/ar\right)h_{ij}^2\right],
\ee

Adding these, the  total contribution of the term containing $E_{ij}E^{ij}$ in Einstein-Hilbert action  to zeroth and the first order in $\mu=MGH$ is:
\begin{itemize}
  \item zeroth order
\be\label{2nd}
\sqrt{\textrm{det}(g_{ij})}\frac{N^{-1}}{a^4(1+MG/2ar)^{8}} \left(E_{ij}E^{ij}\right)=H^2a^3\left[\frac{3}{4}{h}_{ij}^2+\frac{1}{4H^2}\dot{h}_{ij}^2-\frac{1}{H}\dot{h}_{ij}{h}_{ij}\right],
\ee

  \item first order

\be\label{2nd,1}
\sqrt{\textrm{det}(g_{ij})}\frac{N^{-1} \left(E_{ij}E^{ij}\right)}{a^4(1+MG/2ar)^{8}}&=&H^2a^3(MG/ar)\left[\frac{6}{4}{h}_{ij}^2+\frac{1}{H^2}\dot{h}_{ij}^2-\frac{3}{H}\dot{h}_{ij}{h}_{ij}\right] \nonumber\\
&=& \mu Ha^2\frac{1}{r}\left[\frac{6}{4}{h}_{ij}^2+\frac{1}{H^2}\dot{h}_{ij}^2-\frac{3}{H}\dot{h}_{ij}{h}_{ij}\right],
\ee
\end{itemize}

\subsubsection{Term containing $E^2$ }
Starting with
$$
E_{ii}=3Ha^2(1+MG/2ar)^3(1-MG/2ar),
$$
we have
\be\label{c}
E^2=a^{-4}(1+MG/2ar)^{-8}\left[E_{ii}^2-2E_{kk}h_{ij}E_{ij}+2h_{ik}h_{jk}E_{ij}E_{ll}\right] \,.
\ee
Below we calculate the contributions of each of the above three terms:
\ba
\sqrt{\textrm{det}(g_{ij})}\frac{N^{-1} E_{ii}^2}{a^4(1+MG/2ar)^{8}} &=& -\frac{9}{4}H^2 a^3\left(1+MG/ar\right)h_{ij}^2, \\
 \sqrt{\textrm{det}(g_{ij})}\frac{N^{-1} \left(-2E_{kk}h_{ij}E_{ij}\right)}{a^4(1+MG/2ar)^{8}} &=& -6H^2  a^3\left(1+MG/2ar\right)^5\\
&&\times\left[\left(1-MG/2ar\right)h_{ij}^2
+\frac{1}{2H}\left(1+MG/2ar\right)h_{ij}\dot{h}_{ij}\right],\nonumber\\
 \sqrt{\textrm{det}(g_{ij})}\frac{N^{-1} \left(2h_{ik}h_{jk}E_{ij}E_{ll}\right)}{a^4(1+MG/2ar)^{8}} &=&6H^2 a^3\left(1+MG/2ar\right)^5\left[\left(1-MG/2ar\right)h_{ij}^2\right].
\ea
Adding up the above contributions, we have
\be\label{cur3}
 -\sqrt{\textrm{det}(g_{ij})}N^{-1}E^2=H^2 a^3\left[\frac{9}{4}\left(1+MG/ar\right)h_{ij}^2+3(1+3MG/ar)h_{ij}\dot{h}_{ij}\right].
\ee
To first order in $\mu$, we have
\be\label{cur3}
 -\sqrt{\textrm{det}(g_{ij})}N^{-1}E^2=H\mu a^2/r\left[\frac{9}{4}h_{ij}^2+\frac{9}{H}h_{ij}\dot{h}_{ij}\right].
\ee

The total contribution from  Eq. (\ref{2nd,1}) and  Eq. (\ref{cur3}) in  Einstein-Hilbert  action is given by
\begin{itemize}
  \item zeroth order
\be\label{2nd}
\sqrt{\textrm{det}(g_{ij})}\frac{N^{-1}}{a^4(1+MG/2ar)^{8}}\left(E_{ij}E^{ij}-E^2\right)=a^3\frac{1}{4}\dot{h}_{ij}^2
\ee
  \item first order
\be\label{2nd,11}
\sqrt{\textrm{det}(g_{ij})}\frac{N^{-1}}{a^4(1+MG/2ar)^{8}}\left(E_{ij}E^{ij}-E^2\right)=\mu Ha^2(1/r)\left[\frac{-9}{4}h_{ij}^2+\frac{1}{4H^2}\dot{h}_{ij}^2\right] \, .
\ee
\end{itemize}

Combined with the first term of the  Einstein-Hilbert  action given in Eq. (\ref{first0}), the total
Einstein-Hilbert  action to first order in $\mu$ is obtained to be
\begin{eqnarray}\label{E-H 1}
&&{M_P^2\over 2}  \int \sqrt{\det (g_{ij})} N \left(^{3}R+{1\over N^2}\left(E_{ij}E^{ij}-E^2\right)\right)=\nonumber\\
&&~~~~~~~~~~{M_P^2\over 2}\mu\left[4\pi\frac{1}{H}\delta(r)(h_{ij})^2+4\frac{1}{H}\partial_j\partial_k\left(\frac{1}{r}\right)h_{ij}h_{ik}+
\frac{Ha^2}{r}\left(-\frac{9}{4}h_{ij}^2+\frac{1}{4H^2}\dot{h}_{ij}^2\right)\right].
\end{eqnarray}
This gives the four interaction Hamiltonians as presented in Eqs. (\ref{InterH}) - (\ref{InterH4}).


\end{document}